\documentclass[12pt]{article}
\usepackage{graphicx}
\usepackage{amsfonts}
\usepackage{amsmath}
\usepackage{caption}
\usepackage{natbib}
\usepackage{bm}
\usepackage{color}
\usepackage{mathrsfs}
\newtheorem{theorem}{\indent\sc Theorem}

\newtheorem{proposition}{\indent\sc Proposition}

\newcommand{\bbeta}[0]{\boldsymbol \beta}
\newcommand{\bx}[0]{\mathbf x}
\newcommand{\by}[0]{\mathbf y}
\newcommand{\bh}[0]{\mathbf h}
\newcommand{\bu}[0]{\mathbf u}
\newcommand{\bc}[0]{\mathbf c}

\newcommand{\bw}[0]{\mathbf w}
\newcommand{\diag}[0]{\text{diag}}
\newcommand{\Bf}[0]{\mathbf f}
\newcommand{\btheta}[0]{\boldsymbol{ \theta}}

\begin{document}
\title{A Conditional Density Estimation Partition Model Using Logistic Gaussian Processes}
\author{ Richard D.\ Payne, Nilabja Guha, Yu Ding, \& Bani K.\ Mallick}



\maketitle

\begin{abstract}
Conditional density estimation (density regression) estimates the distribution of a response variable $y$ conditional on covariates $\bx$.  Utilizing a partition model framework, a conditional density estimation method is proposed using logistic Gaussian processes.  The partition is created using a Voronoi tessellation and is learned from the data using a reversible jump Markov chain Monte Carlo algorithm.  The Markov chain Monte Carlo algorithm is made possible through a Laplace approximation on the latent variables of the logistic Gaussian process model.  This approximation marginalizes the parameters in each partition element, allowing an efficient search of the posterior distribution of the tessellation.  The method has desirable consistency properties.  In simulation and applications, the model successfully estimates the partition structure and conditional distribution of $y$.
\end{abstract}


\section{Introduction}
Conditional density estimation (sometimes referred to as density regression) is a method used to estimate the conditional distribution of a response variable, $y$, which depends on a vector of covariates, $\bx$.  Many common regression methods are special cases of conditional density estimation.  For instance, the Gaussian regression model assumes that the mean of $y$ changes with $\bx$ with the variance of $y$ being constant over the covariate space.  Conditional density estimation is particularly useful when a parametric form linking the covariate space $\bx$ with $y$ is unknown or violates the assumptions of existing parametric methods.  In its most flexible forms, conditional density estimation can be viewed as a  general nonparametric regression method.

There are a number of existing frequentist approaches to perform density regression including kernel methods \citep{fan1996estimation, fu2011kernel}, spline methods \citep{kooperberg1991study, stone1997polynomial}, and mixtures of experts \citep{jacobs1991adaptive}.  There are also several Bayesian approaches to conditional density estimation.  One of the popular approaches is to use mixture models for the conditional distribution of $p(y \mid \bx)$ and allow the mixing weights as well as the parameters to depend on the covariates \citep{chung2012nonparametric, dunson2008kernel, dunson2007bayesian,griffin2006order}. An alternative approach is to apply the logistic Gaussian process model in the conditional density estimation setting \citep{tokdar2010bayesian}. Latent variable models have been utilized by \citet{kundu2011single} and \citet{bhattacharya2010nonparametric}.  A multivariate spline based method \citep{shen2016adaptive} and an optional Polya tree based method \citep{ma2012coupling} have been recently proposed.  \citet{petralia2013multiscale} perform density regression using a convex combination of dictionary densities using a fixed tree decomposition which scales to accommodate a large number of features.

The purpose of this paper is to provide a novel partition model \citep{holmes2012bayesian,denison2002bayesian} framework to perform density regression using logistic Gaussian processes.  This proposed  method  is intended to describe the distribution of $y$ in a {\it region} of $\bx$.  This method is similar in spirit to the Bayesian classification and regression tree (CART) model \citep{chipman1998bayesian, denison1998bayesian} as a decision tool to understand how and where the density of $y$ changes in different regions of $\bx$.  In fact, we adaptively partition the covariate space and use a logistic Gaussian process model within each region. This method can also be viewed as a non-parametric changepoint analysis in which the changepoints occur in a multi-dimensional covariate space and the distribution of $y$ is not restricted to any specific parametric form. Hence, we can derive a decision rule based on the partitions which makes it practically useful.  For example, a business may be interested in understanding how years of experience and a persuasiveness score influence revenue from individual salespersons (where variability and skewness of revenue may change throughout the covariate space).  The model can identify where major changes in the shape, spread, and/or center of the distribution of revenue occurs among their sales force.  Another example is that of power output from windmills where assuming a fixed error structure of power output, $y$, over the covariate space (wind speed, direction, density, etc.) is not appropriate, nor is the relationship between $y$ and the covariates believed to be linear.  While existing conditional density estimation methodologies allow the estimation of $y$ at a specific point, this methodology allows one to identify where the important changepoints exist, and estimate the density of $y$ in each region of $\bx$. In addition, variable selection is automatically incorporated as irrelevant covariates will not  influence the partitioning in this framework.

Partition models provide an appealing framework to determine the breakpoints in $\bx$ where the density of $y$ changes.   Specifically, we can utilize existing density {\it estimation} methods in constructing a density {\it regression} methodology.  In the simplest cases, partition models fit independent/separate models to each piece of the  partition.  Ideally these partitions are determined using a data-driven approach.  In our current endeavor, we aim to find a data-driven partition of the covariate space $\bx$ and fit independent density estimates for $y$ in each partition.  This type of partition methodology has been used in  spatial applications by \citet{kim2005analyzing} and in disease mapping applications by \citet{denison2001bayesian}. 

We use a logistic Gaussian process model \citep{lenk1988logistic,lenk1991towards} for conditional density estimation within each region of the partition. \citet{tokdar2010bayesian} used similar models, but our method is completely different than their approach. They used the logistic Gaussian process to model the joint distribution of the response $y$ and the covariates $\bx$ and utilized a subspace projection method to reduce the dimension of the covariates. On the other hand, we are fitting univariate logistic Gaussian processes within each region of $\bx$, hence avoiding the curse of dimensionality. Furthermore, in joint modeling approaches, $y$ and $\bx$ are not identified as the response and covariates. Our experience is that the distribution of the response $y$ is highly influenced by the distribution of $\bx$ (especially when the dimension of $\bx$ is high). In addition, the proposed method is highly interpretable as it  explores the relationship between the covariates and the density function in a nonlinear way which may not be available in  projection based methods.  To the authors' knowledge, there is no publicly available code to perform Bayesian conditional density estimation.  Therefore, part of our contribution is providing the code to perform Bayesian density regression in the partition model framework.  This code is publicly available at {\tt https://github.com/gitrichhub/bayes-cde}.

Our proposed method is also different than the existing partition model approaches. The effectiveness of the partition model algorithms depends on explicit marginalization of the model parameters so that an efficient reversible jump Markov chain Monte Carlo (MCMC) algorithm \citep{green1995reversible} can be developed over the number and location of the partitions. Therefore, all the existing partition models are based on the conjugate structure of the likelihood and priors. On the contrary, the logistic Gaussian model does not have a conjugate structure, so explicit marginalization is not possible. Hence, we utilize a Laplace approximation of the logistic Gaussian models \citep{riihimaki2014laplace} to obtain the marginal likelihood in order to develop an efficient reversible jump MCMC algorithm. Furthermore, to the best of our knowledge, none of the existing papers investigated theoretical properties of the partition models. Ours is thus the first paper that considers the posterior consistency in estimating conditional distributions in the partition model framework. Indeed, there are  a few papers which have considered theoretical properties of density regression models using  other modeling frameworks \citep{tokdar2007posterior,pati2013posterior,norets2012bayesian,bhattacharya2010nonparametric}.

In Section 2 we present the conditional density estimation model in a partition framework and provide a reversible jump MCMC algorithm for estimation.  Section 3 discusses some results on consistency.  Section 4 applies the method to both simulated and real datasets and Section 5 concludes.




\section{Bayesian Hierarchical Conditional Density Estimation Partition Model}
\subsection{Modeling the partition structures using a Voronoi tessellation}
\begin{figure}
\centering
\includegraphics[scale=.4]{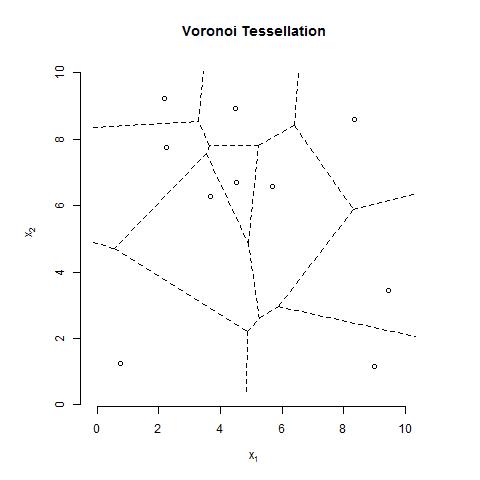}
\caption{A Voronoi tessellation in two dimensions with weight vector $\bw = (.5,.5)$.  The points represent the 10 centers $\bc_1,\ldots,\bc_{10}$ and the dashed lines represent the partition borders.}
\label{fig:voronoi}
\end{figure}

The partition model divides (partitions) the $p$-dimensional covariate space $\mathcal{D}$ into $M$ distinct pieces where $y$ is assumed to independently follow a different density $p_i(\cdot)$ within each partition.   The partitioning of the covariate space is done through a Voronoi tessellation.  The tessellation is defined by $M$ centers $\bc_1,\ldots,\bc_M$ that divide the covariate space into $M$ disjoint regions $R_1,\ldots,R_M$ where $R_i$ consists of all the observed $\bx$ that are closest to center $\bc_i$.  Formally, $R_i = \{\bx \in \mathcal{D} : ||\bx - \bc_i|| < ||\bx- \bc_j|| \ \forall \ i \neq j \}$.  Here, $||\bx|| = ||(x_1,\ldots,x_p)|| = \sum_{i=1}^p w_i x_i^2$ where $\bw$ is a normalized weighting vector ($\sum w_k = 1$) which places different weights on each of the covariates \citep{holmes2012bayesian}.   The weighting provides additional flexibility in the tessellation and also performs variable selection (which will be demonstrated in the examples). Figure~\ref{fig:voronoi} shows an example of a Voronoi tessellation in two dimensions with $\bw = (.5,.5)$.

For simplicity, we assume that the possible centers of the tessellation are restricted to the observed data points $\bx$.  Next, we assign  prior distributions for both the number of centers as well as the center locations.  An intuitive way to express the prior is $p(\bc,M,\bw) = p(\bc \mid M)p(M)p(\bw)$ where

\begin{eqnarray*}
p(M) &=& \text{DU}(M \mid 1,\ldots,M_\text{max}) \\
p(\bc \mid M) &=& \text{DU}\left(\bc \mid 1,\ldots,{n \choose M}\right)\\
p(\bw) &=& \text{Di}(\bw \mid 1,\ldots,1)
\end{eqnarray*}

where DU$(x \mid 1,\ldots,n)$ means discrete uniform on $1,\ldots,n$ and $M_\text{max}$ is the maximum number of allowable centers (a hyperparameter chosen by the user).  The prior on $p(\bc \mid M)$ gives equal weight to all possible combinations of $M$ centers with possible center locations corresponding to the $n$ observed values of the covariates $\bx$.  The vector $\bw$ has the Dirichlet prior which is uniform on the simplex.

\subsection{The data generating model}

In this section, we develop the likelihood function by fitting a logistic Gaussian process model within each element of the partition. We assume that within the $i$th region the data follow a logistic-Gaussian model.  Let $T = \{M,\bc,\bw\}$ denote the tessellation parameters and $\btheta = (\btheta_1,\ldots,\btheta_M)$ denote the set of all other (smoothing) parameters in each partition region. Also let $\by_{i}=(y_{i1},\cdots ,y_{in_{i}})$ and $\bx_{i}$ denote the observed response variable and covariates found in the $i$th region of the partition, $i=1,\ldots ,M$. Then the logistic Gaussian process model within the $i$th region will be
\[
p(y_i) = \frac{\exp(f_i(y_i))}{\int_{\mathcal{V}_i}\exp(f_i(s))ds}
\]

where $f_i(y_i)$ is an unconstrained latent function and $y_{i}$ lies in a finite region $\mathcal{V}_{i}$ of $\mathbb{R}$.  A Gaussian process (GP) prior is placed on $f_{i}$, therefore smoothing can be controlled via the covariance structure. 

We first discretize the region $\mathcal{V}_{i}$ into a regular grid with $r$ subregions centered at $Z_{i} = [z_{i1},\ldots,z_{ir}]^T$.  We then specify the prior on $f_{i}$ by assuming 

\begin{align}
f_i(z_{ij}) &= g(z_{ij}) + \mu_{ij} \nonumber\\
g(z_{ij}) &\sim GP(0,\kappa_{\btheta_{i}}(z_{ij},z_{ij^{'}})) \nonumber \\
\mu_{ij} &= \bh(z_{ij})^T\bbeta_{i} \label{eqn:mu}
\end{align}

 where $\kappa_{\btheta_{i}}(z_{ij},z_{ij^{'}})$ is the covariance function with inputs $z_{ij}$ and $z_{ij^{'}}$ which depends on the smoothing hyperparameter $\btheta_{i}$.  For simplicity in the following exposition, we assume $\btheta_{i}$ is fixed and known, and will discuss its selection later in this section.  By placing a Gaussian prior on $\bbeta_{i} \sim N(\mathbf{b},B)$, $\bbeta_i$ can be integrated out to yield 

\[
f_i(z_{ij}) \sim GP(\bh(z_{ij})^T \mathbf{b},\ \kappa_{\btheta_{i}} (z_{ij},z_{ij^{'}})  + \bh(z_{ij})^TB\bh(z_{ij^{'}})).
\]

Consequently, given a discretization $Z_i$ and covariance parameters $\btheta_i$, the prior on $f_i$ can be expressed as a Gaussian distribution over the latent function values $\Bf_i$

\[
p(\Bf_{i}\ \mid \ Z_{i}, \btheta_{i}) = N(\Bf_{i}\ \mid \ H_{i}\mathbf{b}, K_i + H_{i}BH_{i}^T)
\]

where $\Bf_i = [f_i(z_{ij})]_{j=1}^r$ is a column vector of $r$ latent values associated with each subregion, $K_i = [\kappa_{\btheta}(z_{ij},z_{ik})]_{j,k=1}^r$ and $H_{i} = [\bh(z_{ij})^T]_{j=1}^r$.  Following \citet{riihimaki2014laplace}, we choose $\bh(z_{ij}) = [x,x^2]^T$.

Let $\by_{i}=(y_{i1},\cdots ,y_{in_{i}})$ be the vector of $n_{i}$ responses found in the $i$th partition and $\by_{i}^\star = [y^\star_{ij}]_{j=1}^r$ where $y_{ij}^\star$ is the number of observations of $\by_{i}$ that fall into region $j$ (i.e. closest to $z_{ij}$), then the log-likelihood contributions of the $n_i$ observations is

\[
\log[p(\by_{i} \mid \Bf_{i},Z_{i})] = \by_{i}^{\star T}\Bf_{i} - n_{i}\log\left(\sum_{j=1}^r \exp(\Bf_{ij})\right)
\]

where $\Bf_{ij}$ is the $j$th element of $\Bf_i$. Consequently, the posterior distribution of $\Bf_i$ can be expressed

\[
p(\Bf_{i} \mid \by_i,Z_{i},\btheta_i) = \frac{p(\Bf_i \mid Z_{i},\btheta_i)p(\by_{i} \mid \Bf_{i},Z_{i})}{\int_{\mathcal{V}_{i}} p(\Bf \mid Z_{i},\btheta_{i})p(\by_{i} \mid \Bf,Z_{i}) d\Bf}
\]

and the marginal distribution after integrating out $\Bf_{i}$ is 

\[
p(\by_{i} \mid Z_{i},\btheta_{i}) = \int_{\mathcal{V}_i} p(\Bf_{i} \mid Z_{i},\btheta_{i})p(\by_{i} \mid \Bf_{i},Z_{i}) d\Bf_{i}.
\]

Given a tessellation, $T$, and using the independence assumption among the partition regions, we can express the marginal likelihood as $p(\by \mid \btheta,[Z_i]_{i=1}^M,T)=\prod_{i=1}^{M}{p(\by_{i} \mid \btheta_{i},Z_{i})}$. The conditional distribution of the tessellation parameters, $T$, can be obtained by marginalizing $\Bf_i$ in each region: 

\[
p(T \mid \by, \btheta,[Z_i]_{i=1}^M) \propto p(T)\prod_{i=1}^{M}{p(\by_{i} \mid \btheta_{i},Z_{i},T)}.
\]

This marginalization makes a collapsed MCMC algorithm possible. However, due to lack of conjugacy,  $p(\by_{i} \mid \btheta_{i},Z_{i},T)$ is not available explicitly. Therefore, we use a Laplace approximation \citep{riihimaki2014laplace} to integrate over $\Bf_i$ in each partition element.  The posterior approximation of $\Bf_i$ takes the following form in each partition 

\[
p(\Bf_i \mid \by_i,Z_i,\btheta_i) \approx N(\Bf_i \mid \hat{\Bf}_i,\Sigma_i)
\]

where $\hat{\Bf}_i  = \arg\max_\Bf p(\Bf_i \mid \by_i,Z_i,\btheta_i)$ and $\Sigma_i = ((K_i + H_iBH_i^T)^{-1} + n_i(\diag(\bu_i) - \bu_i\bu_i^T))^{-1}$ with the entries of the vector $\bu_i$ as $u_{ij} = \exp(\hat{\Bf}_{ij})/\sum_{j=1}^r\exp(\hat{\Bf}_{ij})$.  The Laplace  approximation yields an approximate form of the marginal of $p(\by_i \mid  Z_i,\btheta_i)$ which can be used to approximate $P(T \mid \by,\btheta,[Z_i]_{i=1}^M)$ in the MCMC algorithm.   A key strategy in \citet{kim2005analyzing} is to marginalize out all parameters which are specific to individual partitions and develop a collapsed MCMC algorithm to obtain posterior samples of the tessellation, $T$.  Using the posterior samples of $T$, the best partition (in terms of either posterior probability or marginal likelihood of $\by$) is selected and used to make inference and predictions.  

Once a posterior partition is selected, the Laplace approximation makes density estimation within each partition simple and straightforward.  Obtaining $\hat{\Bf}_i$ can generally be done using Newton's method, and is much faster computationally than using MCMC methods to draw from the posterior of $\Bf_i$.   Once $\hat{\Bf}_i$ has been found, the posterior density of $p(y_i)$ is found by simulating from $N(\Bf_i \mid \hat{\Bf}_i,\Sigma_i)$ and applying the log-density transform.  


We now discuss the form of $\kappa_{\btheta_i}(\cdot,\cdot)$ and the selection of $\btheta_i$ in each partition.  For the purposes of this paper, we assume the covariance function $\kappa_{\btheta_i}(\cdot,\cdot)$, which depends on hyperparameters $\btheta_i = (\sigma_i^2,l_i)$, is the stationary squared exponential covariance function 

\[
\kappa_{{\btheta}_i}(z,z^\prime) = \sigma_i^2 \exp\left(-\frac{1}{2l_i^2} (z - z^\prime)^2   \right)
\]

where $\sigma_i^2$ is the magnitude hyperparameter and $l_i$ is a length-scale hyperparameter which together govern the smoothness properties of $f_i$.  We place a weakly informative half Student-$t$ distribution with one degree of freedom and a variance equal to 10 for the magnitude parameter and the same prior with a variance of 1 for the length-scale hyperparameter \citep{riihimaki2014laplace}. We then obtain the maximum a posteriori (MAP) estimate of the posterior mode of $\btheta_i$ using the Broyden-Fletcher-Goldfarb-Shanno (BFGS) quasi-Newton algorithm and set $\btheta_i$ equal to this value.  This approach to selecting and fixing $\btheta_i$ can be viewed as an empirical Bayes prior.

\subsection{Markov Chain Monte Carlo algorithm}

The posterior distribution of the tessellation structure, $T$, does not have an explicit form, thus obtaining posterior draws of the tessellation proceeds using a reversible jump MCMC algorithm \citep{green1995reversible}.  A reversible jump MCMC algorithm is necessary due to the varying dimension of the parameter space of the tessellation structure.   The algorithm to explore this varying dimensional parameter space proceeds as follows:

\begin{enumerate}
\item Initialize the tessellation structure by choosing one observation's covariate vector as the initial center of the tessellation and choosing a starting weight $\bw$ (a simple default is to set $\bw = (p^{-1},\ldots,p^{-1})$).
\item When $1 < M < M_\text{max}$, with probability 1/4, either add (birth step), delete (death step), or move (moving step) a tessellation center; otherwise, modify the weight vector $\bw$.  A new center is chosen uniformly from the locations (observed covariate vectors) that are not currently a tessellation center, a center is deleted by randomly choosing an existing center and removing it.  Moving a tessellation center entails uniformly choosing one existing center and randomly moving it to another observation's covariate vector which is not currently a tessellation center.  To change the weight vector, a new weight vector $\bw^{(p)}$ is proposed by drawing from $q(\bw^{(p)} \mid \bw) \equiv \text{ Dirichlet}(dw_1,\ldots,dw_p)$, where $\bw = (w_1,\ldots,w_p)$ is the vector of the current tessellation weights in the MCMC chain and $d>0$ is a tuning parameter which determines the variance of the proposal distribution $q(\cdot \mid \cdot)$.  Accept the proposed change to the tessellation with probability $\alpha$, defined below.

\item Repeat step 2 until the MCMC chain has converged and enough samples have been generated.  Discard burn-in period.

\end{enumerate}

Let $T_p$, $T_c$ be the proposed and current tessellation structures with $M_p$ and $M$ partitions, respectively.  We accept the birth, move, death, or change in weight step using the acceptance probability

\begin{equation} \label{eq:ratio}
\alpha = \min \left(1, \frac{q(\bw \mid \bw^{(p)}) \prod_{i=1}^{M_p}  p(\by_i^\star \mid T_p,Z,\btheta_i)}{q(\bw^{(p)} \mid \bw) \prod_{i=1}^M p(\by_i^\star \mid T_c,Z,\btheta_i)} \right).
\end{equation}

It is important to note that for most non-boundary cases, the prior on the tessellation structure $P(T_p)$ and $P(T_c)$ does not appear in $\alpha$ due to cancellations with itself and/or the proposal distribution for the birth, death, move, and changing weight steps in the reversible jump MCMC algorithm. When $M$ is at or near the boundary, however, adjustments to $\alpha$ need to be made in order to maintain the reversibility of the MCMC chain. When $M = 1$ and a birth step is proposed,  we must multiply the ratio in \eqref{eq:ratio} by 3/4.  When $M=2$ and we propose a death step, the ratio must be multiplied by 4/3.  The reverse must be applied when $M=M_\text{max},\text{ } M_\text{max} - 1$.  Note that if the weight vector $\bw$ remains unchanged (i.e. birth, death, and move steps), the ratio involving $q(\cdot \mid \cdot)$ will simply be 1.

For a given tessellation drawn from the posterior distribution of $T$, computing the density in each partition of the tessellation is done by simply computing a univariate density estimation using the logistic Gaussian process model.  One can then choose the tessellation (partition) with highest posterior probability or marginal likelihood of $y$ and use this as the final density regression model.  


\section{Some Results on Convergence}
In this section, we show the consistency of the proposed conditional density estimation method. If the proposed model is true, then as $n$, the number of points goes to infinity,  the posterior density concentrates near a small total variation neighborhood around the true density. We first state the result for the Euclidean norm, denoted by the metric $d$. For a general weighted norm, the result is extended in the appendix. 

To prove this result, we first show that the partitions formed by a Voronoi tessellation can adequately approximate the true partition. Then, we show that we have sufficient prior probability for the approximating partition and  around any small neighborhood of the true Gaussian process  path in supremum norm. Finally, if we have sufficient prior mass around the true density, the likelihood pulls the   posterior density towards the data generating density under the true model.

Let ${\mathbf c}_1,{\mathbf c}_2,\dots,{\mathbf c}_m$ be the centers of some Voronoi tessellation and $R_1,R_2,\dots,$ $R_m$ be the corresponding Voronoi regions in $\Omega$, a  subset of $\mathbb{R}^d$ with associated  Lebesgue measure $\mathscr{L}$. Let $V_1,\dots,V_k$ be any given partition of $\Omega$. We assume that each region, $V_i$, is a finite union of rectangular regions. Our result holds for a general region approximated by a finite union of rectangles. We first prove that any aforementioned  region $V_i$, can be approximated by the regions of a Voronoi tessellation. 
\begin{proposition}
	Given $\epsilon_1>0$ there exists $M$ and ${\mathbf c}_1,\dots, {\mathbf c}_M$ and a partition $J_1,\dots,J_k$ of $\{1,\dots,M\}$ such that $U_l=\cup_{i\in J_l}R_i$ and $\sum_{l=1}^k\mathscr{L}(U_l \Delta V_l)\leq \epsilon_1$, where $\Delta$  denotes the symmetric differences of sets.
	\label{vor1}
\end{proposition}

In our proposed method, we use the observed values of the covariates for the centers of the tessellation.  
Next, we show that a small perturbation of ${\mathbf c}_1,\dots,{\mathbf c}_M$ from Proposition \ref{vor1} does not change the partition dramatically and provides an approximation for regions $V_1,\dots,V_k$. Then, we show that any small neighborhood of ${\mathbf c}_1,{\mathbf c}_2,\dots,{\mathbf c}_M$ contains observed covariates with probability 1 as $n$ goes to infinity. For that we assume that the probability measure on ${\bf x}$, $\tilde{H}(\cdot)$ has a strictly positive, bounded density function.   This conclusion implies that we can use covariate points as the centers for the proposed Voronoi tessellation to approximate true partition of the covariate space. We summarize these two results in the form of two following propositions.

\begin{proposition}
	Given $\epsilon_1>0$, ${\mathbf c}_1,\dots,{\mathbf c}_M$ and $R_1,\dots,R_M$ from Proposition \ref{vor1}, we can have $\delta>0$  and Voronoi centers $\mathbf{c}'_1,\dots,\mathbf{c}'_M$ and corresponding $R'_1,\dots,R'_M$ such that  if $d(\mathbf{c}_i,\mathbf{c}'_i)<\delta$ then $\sum_{l=1}^k\mathscr{L}(U'_l \Delta V_l)\leq 2\epsilon_1$, where $U'_l=\cup_{i\in J_l}R'_i$ and  $d$ denotes the distance under Euclidean norm. 	\label{vor2}
\end{proposition}

\begin{proposition}
	Under the setup of Proposition \ref{vor1}, as  $n \rightarrow \infty$, we observe $ \ \mathbf{x}_j \in (\mathbf{c}_j\pm \frac{\delta}{2})$ for all $ 1\leq j\leq M$ with probability 1.
	\label{vor3}
\end{proposition}
Let 
\begin{eqnarray}
g_\mathbf{x}(y)&=&e^{\mu_j+\eta_{j}(y)}{\mathbf 1}_{\mathbf{x} \in U_j}\nonumber \\
f(y \mid {\bf x})&\propto& g_\mathbf{x}(y)
\label{density}
\end{eqnarray}
$\mu_j$ is the mean function given in \eqref{eqn:mu}  and $f^*$ denote the true density.   Let $|\eta_j^*(y)|<k_0, k_0>1$, corresponding to $f^*$. We also assume $\eta_j^*$ to be smooth. Let $\epsilon_2<2k_0\epsilon_1$.  We drop the subscript $\mathbf{x}$  from $f$  for notational convenience. Let $V_1^*,\dots,V_k^*$ be the true underlying partition of $\Omega$ and from Proposition \ref{vor1} there exists $U_1^*,\dots,U_k^*$ from the Voronoi approximation. Let $|\mu_j^*|<k_0$ be the true mean function in the $j$th region. Consider the   following  neighborhood in supremum norm ($\|\|_\infty$) \[N_1=\{\|\eta_j(y)-\eta_j^*(y)\|_\infty<\epsilon_2 \text{ in }{\bf x} \in U^*_j \cap V^*_j \text{ and } \|\eta_j(y)\|_\infty < k_0 \text{ for } {\bf x}\in U_j^* \Delta V_j^*  \}, \]
\[N_2=\{\|\mu_j(y)-\mu_j^*(y)\|_\infty<\epsilon_2 \text{ in }{\bf  x} \in U^*_j \cap V^*_j \text{ and } \|\mu_j(y)\|_\infty < k_0 \text{ for } {\bf x}\in U_j^* \Delta V_j^*  \}.\]
\begin{proposition}
	For $\{\mu_j,\eta_j\}$ pairs such that $\mu_j \in N_2$  and   $\eta_j \in N_1$, $\forall j$;  we have  $\int|f(y)-f^*(y)|<k\epsilon_2$, for some $k>0$. 
	\label{neighbor1}
\end{proposition}

The covariance kernel between points $s$ and $t$ can be written as $K(s,t)=\sigma^2K_0(\frac{1}{l_i}s,\frac{1}{l_i}t)$ for the $i$th region  $R_i$, where $K_0(
\cdot,\cdot)$ is a smooth kernel. 
We assume the following,
\begin{itemize}
	\item[] $\mathbf{A1}: \log(\text{max }\{\Pi(\sigma>\lambda_n),\Pi(\frac{1}{l_i}>\nu_n)\})=O(-n)$.
	\item[] $\mathbf{A2}: M_n^2\lambda_n^{-2}\nu_n^{-2\alpha}/n \rightarrow \infty$.
	\item[]$\mathbf{A3}: {M_n}^{1/\alpha}=O(n^\gamma),0<\gamma<1$.
\end{itemize}
Here, $M_n$ is of polynomial order of $n$, $\lambda_n$ and $\nu_n$ are two sequences of constants, and $\alpha\geq 1$, an integer.  

Let $\eta()$ be any Gaussian process path, under the smoothness of the covariance kernel the paths are smooth and the derivative process is again a Gaussian process. 
For any density based on $m\leq M$ partitions we have an $m$-dimensional  product function space.  We construct sieves on the function space where the probability outside the sieves decreases exponentially with $n$ and establish an entropy bound for the sieves. We use this construction to prove our following  convergence result. 

%
%
%
%

\begin{theorem}
	Let $U_{\epsilon'}=\{f(y): \int |f(y)-f^*(y)|dy<\epsilon'\}$. Then, under $\mathbf{A1}-\mathbf{A3}$ and  log Gaussian process prior  and model \eqref{density}, $\Pi(U_{\epsilon'} \mid \cdot)\rightarrow 1 $ with probability one, as $n$ the number of observations goes to infinity.	\label{thm1} 
\end{theorem}

Even though the main results focus on the neighborhood of the estimated density, the prior favors smaller partitions. Heuristically, if the true partition is further partitioned into smaller partitions, then the true likelihood remains the same over the smaller partitions, but the prior puts $O(n^{-m})$ weight on a partition with $m$ centers. Hence, extra sub-partitions will reduce the posterior probability. Therefore, we should have higher posterior probability for the smaller number of Voronoi centers, as long as it can capture the true  data generating partition. We can use a prior satisfying $\mathbf{A1}-\mathbf{A3}$, or truncate the hyperparameters at $\lambda_n$  and $\nu_n$. 

It is important to note that the application of this model differs from the theory in several small ways.  First, the applied methodology uses two approximations: the discretized version of $f_i$ in each partition, and the Laplace approximation of the marginal of $\by$.  The theory, of course, is not based on these practical approximations, but the results must necessarily depend on a reasonable approximation.  A measure of the closeness of these approximations to the true underlying model is not undertaken in the present paper, but empirical results of \citet{riihimaki2014laplace} indicate that these approximations are reasonable in practice for density estimation.
	
The theory presented provides consistency statements on models which are more general than the applied model.  The applied model assumes that the true partition structure is a Voronoi tessellation whereas the theory allows for any partition whose elements are made up of unions of rectangular regions.  Modeling this more general structure is beyond the current endeavor of this paper, but opens the door for an even more flexible modeling approach where assuming the true partition structure is a single Voronoi tessellation is too restrictive.

\section{Simulations \& Applications}
\subsection{Preliminaries}
We implement the partition model on several simulated and real datasets.  Since the goal of this methodology is to provide insight as to how the density of $y$ changes with covariates $\bx$, the maximum number of partitions for these simulations and applications is capped at 10 to aid in interpretation.  Each MCMC chain was run for 10,000 iterations with a burn in period between 1,000 and 2,000 iterations.  The data are also centered and scaled to provide numerical stability and allow greater interpretability of the weight vector, $\bw$, which performs variable selection on the standardized covariates. 

The code to run this algorithm was written in Matlab and utilizes portions of the excellent code developed by \citet{vanhatalo2013gpstuff}.  A basic function to carry out these types of models is available at {\tt https://github.com/gitrichhub/bayes-cde}.

\subsection{One partition}
We begin with a very simple simulation with $n=1,000$ data points generated from the following model: 

\[
y \sim N(5,.5^2),\text{ }  x_1 \sim N(0,1),\ x_2 \sim N(0,5^2)
\]

Note that in this case the simulated data $\by$ have no relationship with $\bx$, and therefore the method should favor no splitting of the predictor space (i.e. $M=1$).   The MCMC algorithm assigned a 99.85\% posterior probability of no splitting of the covariate space and .15\% probability of having a partition with two regions, indicating that the model appropriately identified the appropriate relationship between $y$ and $x_1,\ x_2$. 

\subsection{Piecewise regression}

\begin{figure}[t]
	\centering
	\includegraphics[scale=.2]{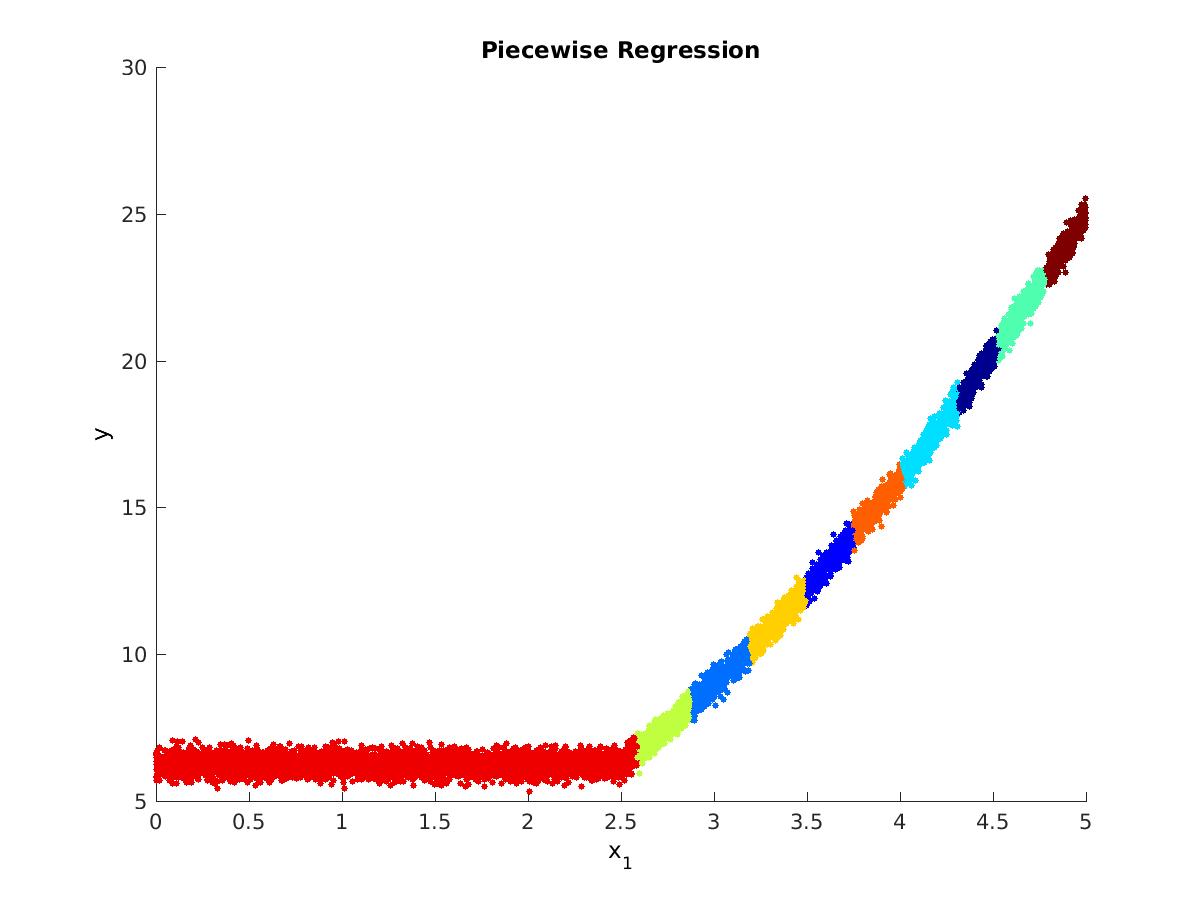}
	\caption{A plot of the response variable $y$ on predictor $x_1$.  The colors of the points represent distinct regions from the posterior partition with the highest marginal likelihood of $y$.}
	\label{fig:piecewise}
\end{figure}

Data ($n=10,000$) was generated by simulating $x_1 \sim U(0,5),\ x_2 \sim N(3,2^2)$, and $y \sim N(f(x_1),.25^2)$ where 

\[
f(x) =  \begin{cases} 
2.5^2 & x < 2.5 \\
x^2 & x \geq 2.5
\end{cases}
\]

Note in this case we have introduced a covariate, $x_2$, which does not have a relationship with $y$.  As expected, the weight for variable $x_2$ was extremely small (less than $1.5*10^{-11}$) for the entire MCMC chain after burn in, indicating the weight vector works well as a variable selector in this framework.  

Figure~\ref{fig:piecewise} plots $y$ against $x_1$.  Colors represent distinct regions from the posterior tessellation with the highest marginal likelihood of $y$.  The method  successfully identified that the distribution of $y$ is the same for $x < 2.5$ and successfully captured changes in the mean function $f(x)$ for $x > 2.5$.  If the maximum number of allowed partitions is increased to 100, then 65-70 partitions are chosen by the model resulting in a much finer partition over $x > 2.5$.

\subsection{Bivariate surface}

\begin{figure}
	\centering
	\includegraphics[scale=.25]{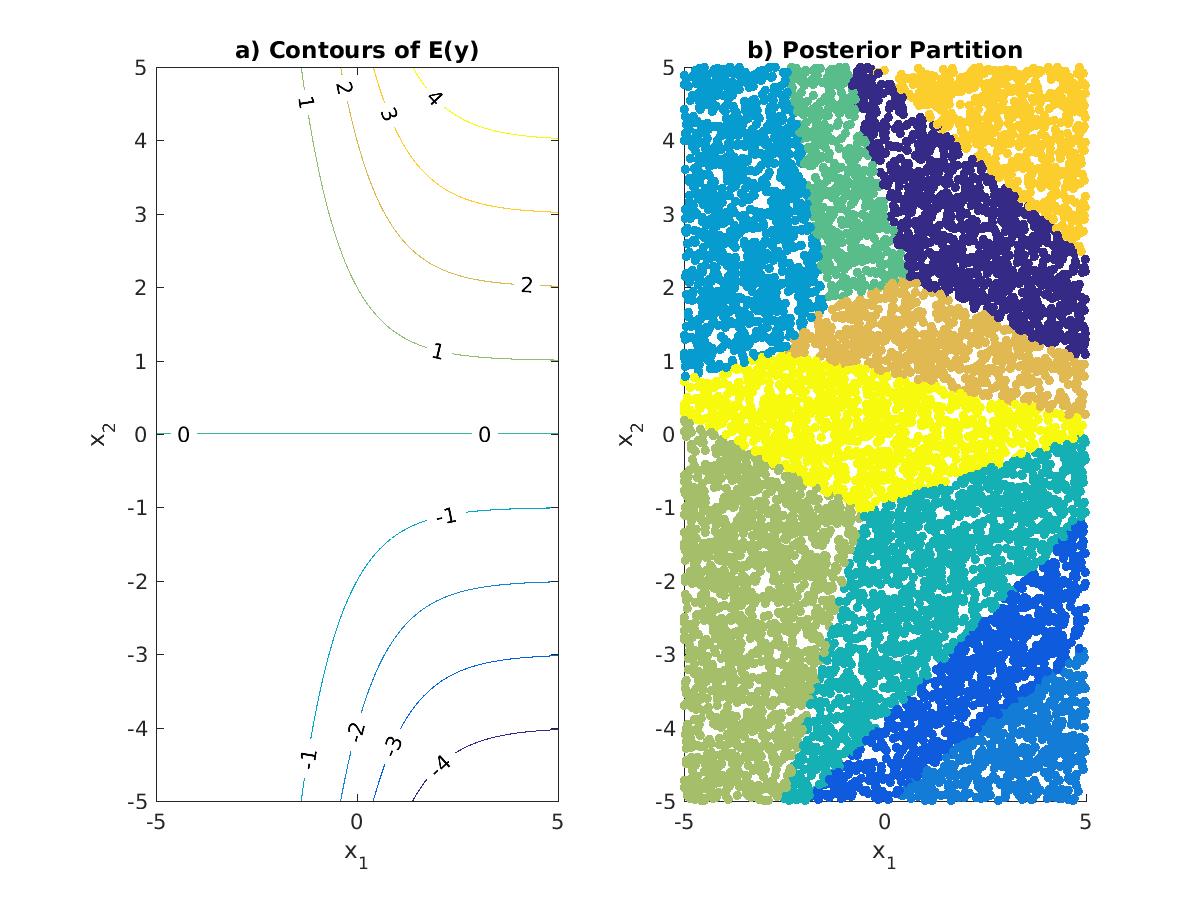}
	\caption{a) The contours of the mean function of $y$. b) The selected posterior partition of $x_1,\ x_2$ with highest marginal probability of $y$. The colors of the points represent distinct partitions.}
	\label{fig:bivariate}
\end{figure}

In this simulation we move to the case where the mean of $y$ is a function of two covariates.  Data ($n=10,000$) was generated from the following model:

\begin{eqnarray*}
x_1,\ x_2 &\sim& U(0,5)\\
y &\sim& N([1 + e^{-x_1}]^{-1}x_2,\ .25^2)
\end{eqnarray*}

For a given value of $x_1$, the mean function of $y$ is a straight line in $x_2$ with slope $[1 + e^{-x_1}]^{-1}$.  Small values of $x_1$ correspond to a slope in the $x_2$ direction near 0, and large values limit to a slope of 1 in the $x_2$ direction.  

The contours of the mean function of $y$ are plotted in panel A of Figure~\ref{fig:bivariate} and the posterior partition (with highest marginal likelihood of $y$) for $x_1$ and $x_2$ is plotted in panel B, with colors indicating distinct regions of the tessellation. The tessellation successfully captures the general features of the mean function of $y$.  When $x_1$ is small, there is little change in the mean of $y$, and the tessellation structure assigns large areas of $\bx$ to the same region.  As $x_1$ increases, we see that a finer partition is induced as the mean structure changes more rapidly over $x_2$.

\subsection{Changing parametric form}

\begin{figure}
	\centering
	\includegraphics[scale=.2]{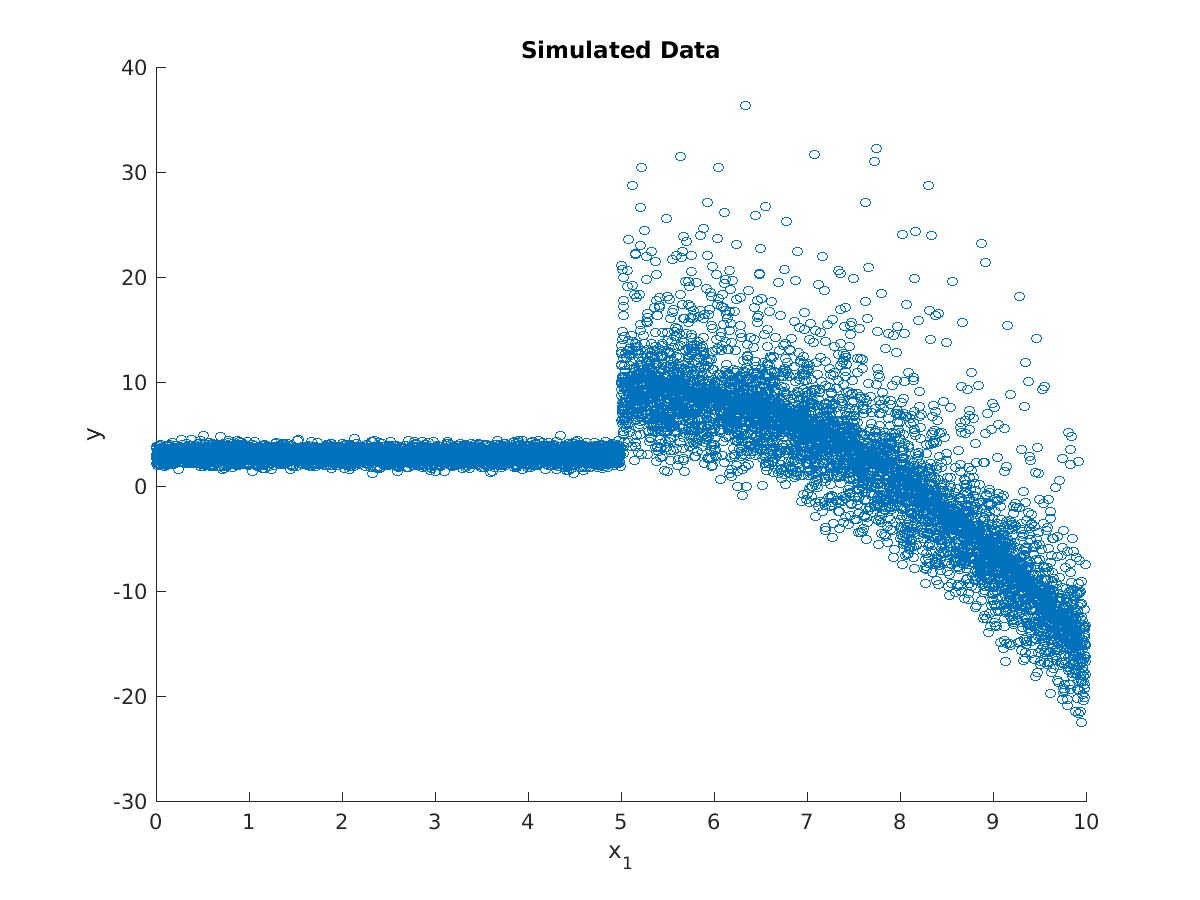}
	\caption{The response variable $y$ against $x_1$, the variable which governs the mean of $y$.}
	\label{fig:complexyx}
\end{figure}

The previous simulated examples have dealt with the case where the distribution of $y$ is related to $x$ only through the mean of $y$.  This example allows the mean function, parametric form, and parameters of the distribution of $y$ to vary with $\bx$.  Specifically, 

\[
y\ \mid \ x_1,x_2 \sim \begin{cases}
N(3,.5^2)&\ x_1 < 5\\ 
-(x_1 - .5)^2 + Z_{x_2}&\ x_1 \geq 5 
\end{cases}
\]
 where $Z_{x_2} \sim \text{Gamma}(2,x_2)$ with a location shift such that $EZ_{x_2} = 0$.  Thus, for a given value of $x_1 \geq 5$, the distribution of $y$ is a mixture of Gamma random variables.  In this simulation, $x_1 \sim U(0,10),\ x_2 \sim U(0,5),\text{ and  }x_3 \sim N(0,5^2)$.  As before, $n=10,000$ and the maximum number of partitions was set to 10.  A plot of $y$ against $x_1$ is given in Figure~\ref{fig:complexyx}.

 Again, we have a variable, $x_3$, which is not related $y$.  The weights of the three variables in the posterior partition with highest marginal likelihood of $y$ reflect the fact that $x_1$ is the dominant predictor with $x_2$ playing a much smaller role and $x_3$ playing essentially no role ($\bw = (.9814,.0185,.0001)$). 

\begin{figure}
	\centering
	\includegraphics[scale=.25]{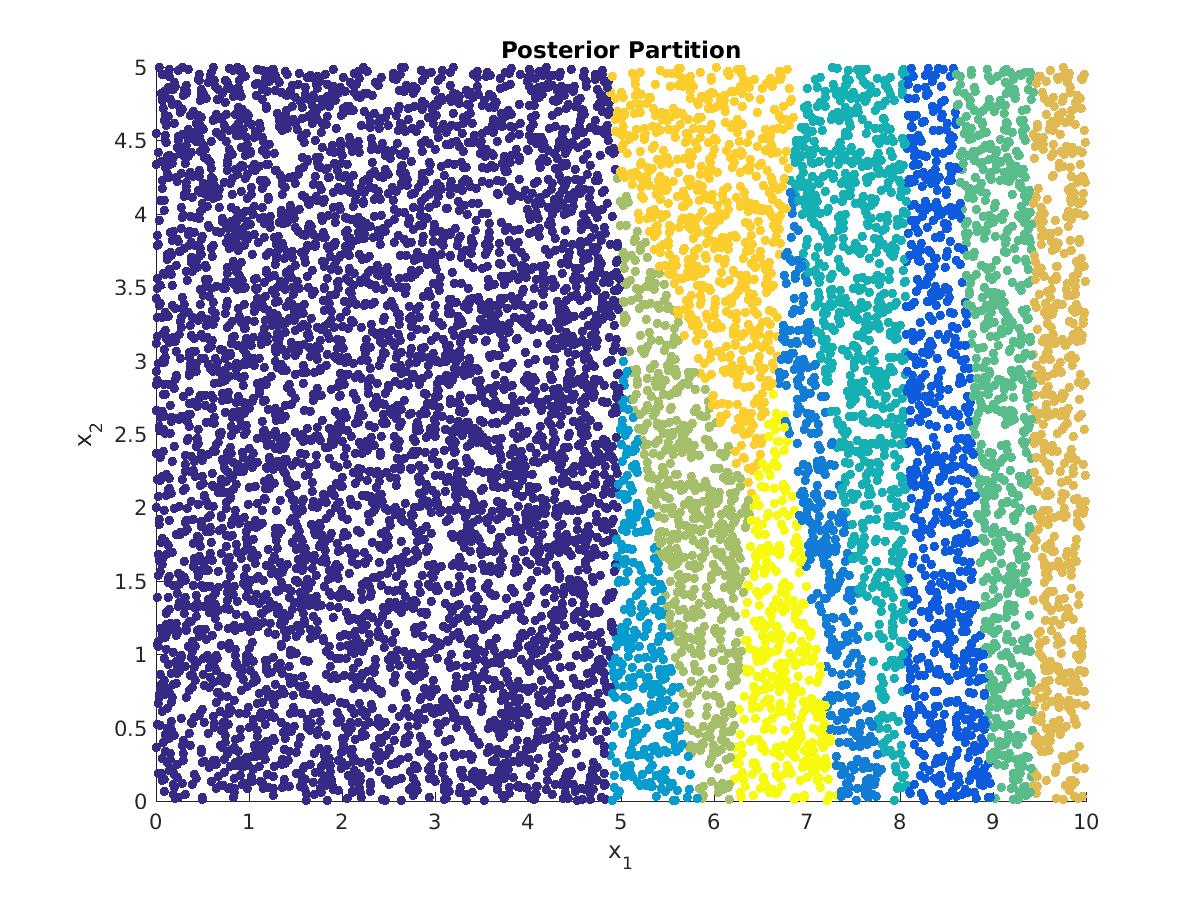}
	\caption{Posterior partition with the highest marginal probability of $y$.  Colors indicate distinct partitions.}
	\label{fig:complex}
\end{figure}

Figure~\ref{fig:complex} shows the posterior partition with highest marginal probability of $y$.  The method successfully identifies the partition for $x_1 < 5$.  Interestingly, when the mean of $y$ is changing slowly (when $x_1$ is just above 5), the model identifies $x_2$ as having an important effect (i.e. for a given $x_1$ value between 5 and 7, the partition changes as a function of $x_2$).  When $x_1$ is larger (and the mean of $y$ is changing faster), $x_2$'s effect is considered negligible compared to that of $x_1$ (i.e. the partition doesn't change over $x_2$ when $x_1$ is large).

\subsection{Melbourne temperature data}

\citet{hyndman1996estimating} introduced several interesting datasets which are well suited for conditional density estimation methods.  One of these is the Melbourne temperature dataset which records each day's maximum temperature (in Celsius) between 1981 and 1990 in Melbourne, Australia.  Interestingly, when each day's maximum temperature is plotted against the previous day's maximum temperature, the temperatures fork as the temperature on the x-axis increases, making it unsuitable for usual regression or other non-parametric regression techniques which generally assume a normally distributed error.

\begin{figure}
	\centering
	\includegraphics[scale=.25]{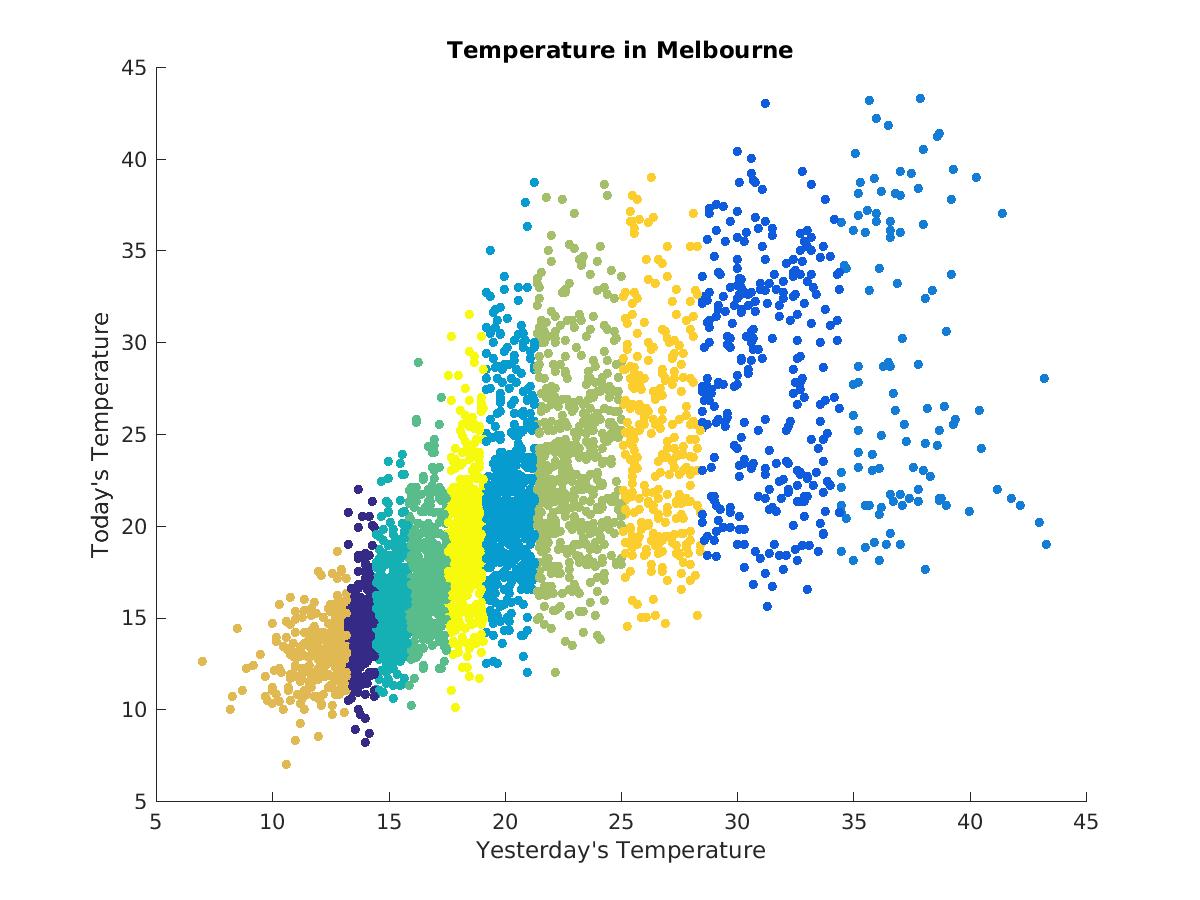}
	\caption{A plot of today's high temperature vs.\ yesterday's high temperature in Melbourne, Australia, 1981-1990.  Colors indicate the partition elements of the tessellation which maximizes the marginal probability of $y$.}
	\label{fig:tempprt}
\end{figure}

\begin{figure}
	\centering
	\includegraphics[scale=.25]{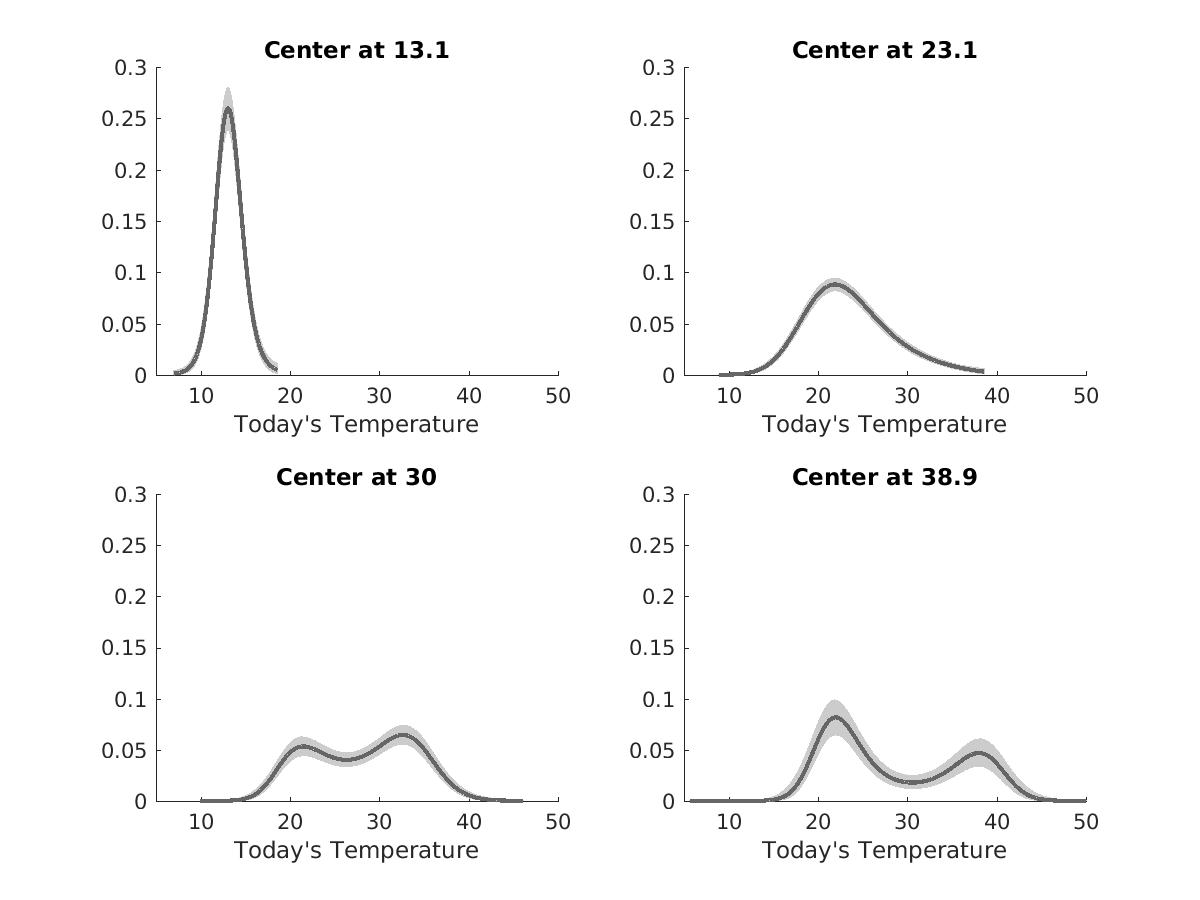}
	\caption{Posterior density estimates (and shaded 90\% credible intervals) of today's temperature in partitions with various centers.  As yesterday's temperature increases (based on the centers), today's temperature transitions to a bimodal form.}
	\label{fig:tempprtind}
\end{figure}

The partition model was applied to this dataset, and the most common partition is shown in Figure~\ref{fig:tempprt}.  Figure~\ref{fig:tempprtind} shows the estimated density of $y$ in several of the partitions.  These densities reveal a steady progression towards bimodality in the distribution of today's temperature as yesterday's temperature increases.  \citet{hyndman1996estimating} explains that this bimodality is a consequence of high pressure systems passing over the city, which are sometimes followed by cold fronts during hotter months (resulting in significantly cooler days following hot days).  Thus hot days are followed by hot days if there is no cold front or significantly cooler days when a cold front passes over Melbourne.

\subsection{Dow Jones changepoint analysis}
\begin{figure}
	\centering
	\includegraphics[scale=.18]{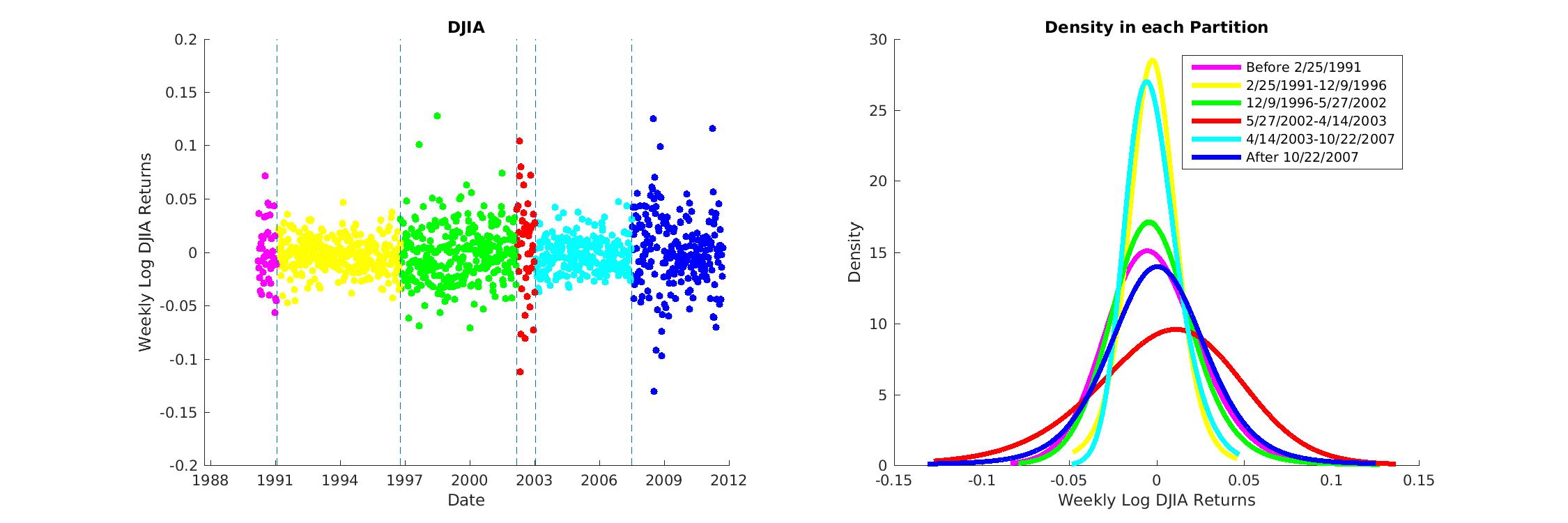}
	\caption{The figure on the left shows the weekly log returns from the DJIA index over time.  Vertical lines represent estimated changepoints (boundaries of the tessellation) and the colors indicate the separate regions from the tessellation with the highest marginal probability of $y$.  The figure on the right panel shows the mean posterior estimate of the distribution of weekly log DJIA returns in each partition.}
	\label{fig:djia}
\end{figure}

The partition model framework can also be used to perform changepoint analyses.  \citet{james2013ecp} perform a changepoint analysis on the weekly log returns of the Dow Jones Industrial Average (DJIA) index.  Applying our partition model framework to this dataset yields 6 partitions (5 changepoints).  The returns are plotted in Figure~\ref{fig:djia} with colors representing the posterior tessellation with highest marginal likelihood.  \citet{james2013ecp} identify 4 changepoints in their analysis, of which 3 correspond closely with the results from our model (12/9/1996, 4/14/2003, and 10/22/2007).  The other two changepoints from our model are 2/25/1991 and 5/27/2002.

The selected changepoints have some interesting relationships with world market conditions.  The changepoint in February 1991 corresponds to the Japanese asset price bubble. The changepoint identified in May of 2002 corresponds to the U.S. stock market downturn of 2002. The changepoint in April of 2003 occurs during the U.S. invasion of Iraq, and the October 2007 changepoint is near the time of sub-prime mortgage crisis.  The right panel of Figure~\ref{fig:djia} shows the posterior mean densities of the returns in each partition.  There appears to be three groups with similar periods of volatility: February 1991 to December 1996 and April 2003 to October 2007 appear to be marked by lower levels of volatility whereas the period between May 2002 and April 2003 is marked by higher volatility.  The remaining time periods appear to have similar levels of volatility.

One major advantage of this method in changepoint analysis is its ability to estimate the density in each region of the partition.  We are not limited to assuming any parametric form of the density in each region, and therefore our the estimates of the density of the returns are not restricted to Gaussian or symmetric heavy tailed distributions.  Indeed, the period between 5/27/2002 and 4/14/2003 is slightly left skewed.

\subsection{Windmill data}
Conditional density estimation is particularly useful when it is unclear how the density of $y$ changes with respect to the predictors, $\bx$.  The exact relationship of electrical power output in windmills with various covariates (wind speed, wind direction, air density, wind sheer, \& turbulence intensity) is unknown, and does not appear to follow any known parametric form.  


We analyze a wind turbine dataset in order to predict the average power output given a set of predictors.  Using a random subsample of 10,000 observations from a larger wind turbine dataset, the partition model was fit to the data using 5 covariates (wind speed, wind direction, air density, wind sheer, \& turbulence intensity).  Density estimates were fit in each region of the tessellation with the highest marginal likelihood.

\begin{figure}
	\centering
	\includegraphics[scale=.25]{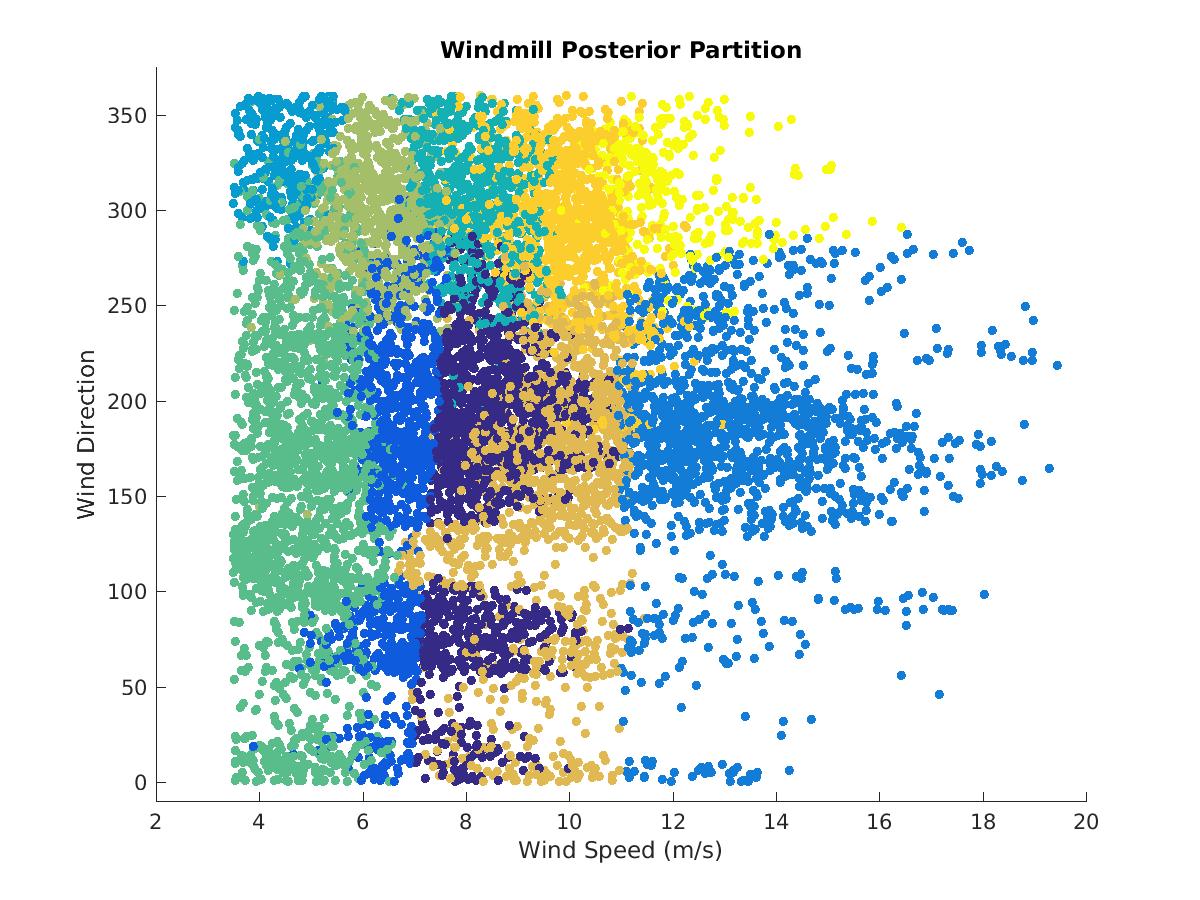}
	\caption{Wind direction plotted against wind speed with colors denoting the partitions of the tessellation with the highest marginal probability of $y$.  Wind speed and direction describe much of the partition structure. Overlaps and blurred edges of the partitions indicate the role of other covariates in determining the partition structure.}
	\label{fig:windpost}
\end{figure}


Figure~\ref{fig:windpost} shows a plot of the posterior partition structure with two of the covariates, wind speed and direction.  From the plot, it is easy to see the importance of wind speed and direction in determining the overall partition structure.  The overlap in regions in this two-dimensional view of the partition indicates the effect of other covariates in determining the 5-dimensional partition structure.

\begin{figure}
	\centering
	\includegraphics[scale=.2]{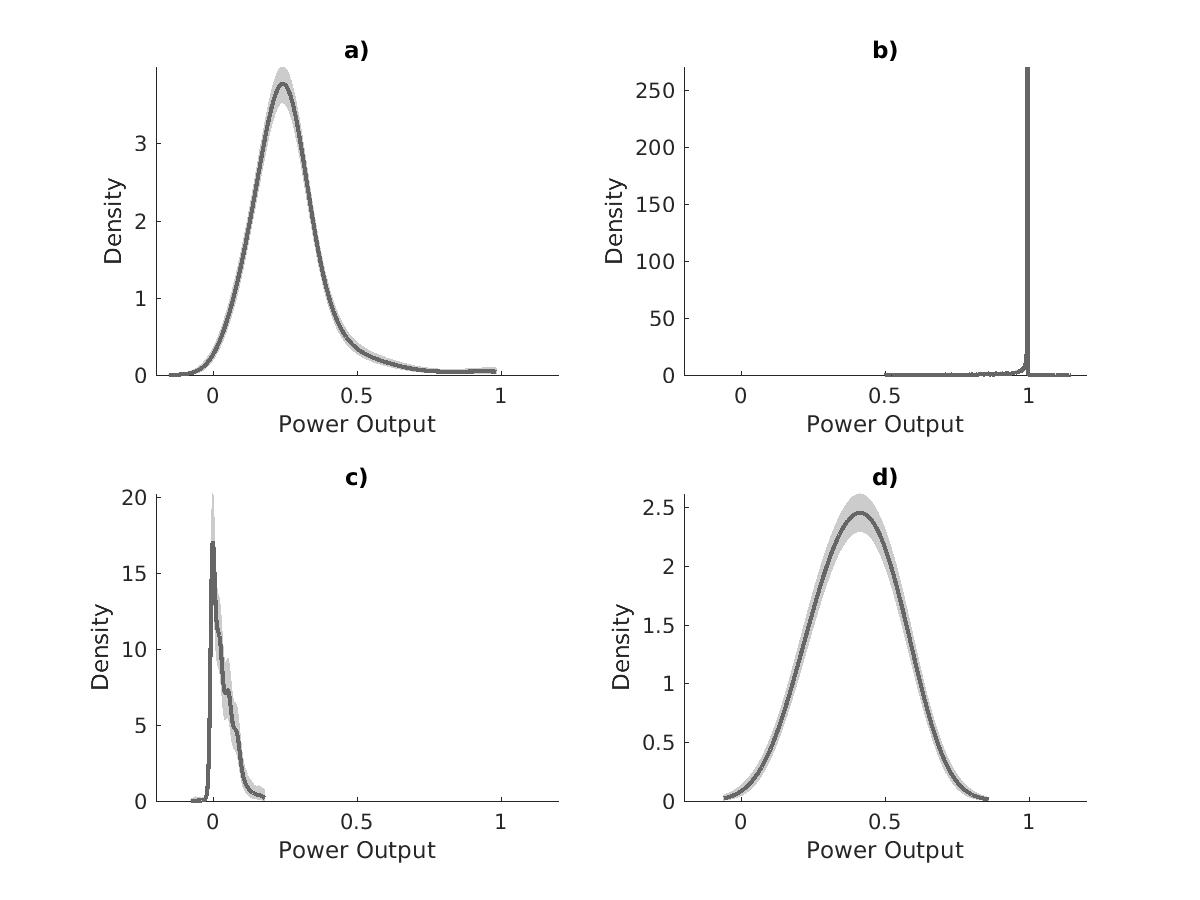}
	\caption{These four posterior densities (and 90\% credible intervals) show the dramatic changes in location, spread, and shape of the density of normalized power output over 4 of the 10 partitions in the final posterior tessellation.}
	\label{fig:windmultprt}
\end{figure}

Figure~\ref{fig:windmultprt} shows posterior densities (with 90\% confidence bands) in four regions of the tessellation.  Note how the distribution of power output changes dramatically throughout the covariate space, demonstrating the need for a density regression technique to more accurately determine the density of $y$ in various regions of $\bx$.


\section{Conclusion}
The combination of logistic Gaussian process density estimation with partition modeling provides an excellent framework for determining how and where the density of $y$ changes over the covariate space.  Furthermore, it has desirable consistency properties.  The applications of this model will help analysts determine which variables are important in predicting the density of $y$, as well as better estimate the density of $y$ over the covariate space.  This flexible method will help give greater insight into datasets in which the relationship between the density of $y$ and $\bx$ is unknown or difficult to formulate parametrically.  As part of our contribution, we provide the first (to our knowledge) publicly available software to perform Bayesian conditional density estimation.  The code is publicly available at  {\tt https://github.com/gitrichhub/bayes-cde}.

 \section{Appendix}

 We show the result from  Proposition \ref{vor1} and \ref{vor2} for  any rectangular region in $d$-dimensional covariate space. We show that any rectangular region can be approximated up to any accuracy by a union of Voronoi regions.  The result is shown using a grid-based  tessellation construction. For general regions, we extend the result thereafter. We then prove our main theorem and address the Voronoi region construction issue under the weighted norm. Later, we address a special case where the true partition in the covariate space is a Voronoi  tessellation. 
 \subsection{Proof of Proposition \ref{vor1}}

 Let $\xi>0$ be any generic small constant. Let $R$ be any $d$-dimensional rectangle with minimum edge length $l$ and maximum edge length $L$. Assume $l>8d\xi$. We construct two outer rectangles $R^{(1)},R^{(2)}$ with edge length $L_i+4d\xi$ and $L_i+8d\xi$, and two inner rectangles $R_{(1)}, R_{(2)}$ with edge lengths $L_i-4d\xi$ and $L_i-8d\xi$, where $L_i$ is the length of $i$th edge. 
 
 The following gives the explicit description of $R^{(1)}$ and $R_{(1)}$. Rectangles $R^{(2)}$ and $R_{(2)}$ can be constructed similarly. 
 In particular, let $\overline{A_iB_i}$ be the edge with length $L_i$ and  $O$ be the center of mass of the rectangle. Let,   $\overline{A_i'B_i'}$ be the  edge of the outer rectangle with same center of mass $O$ and length $L_i+4d\xi$, and $\overline{a_i'b_i'}$ is the  edge of inner rectangle with center of mass $O$ and length $L_i-4d\xi$. Here, $A_i'$, $a_i'$ lie on the half-line  $\overrightarrow{OA_i}$, and $B_i'$, $b_i'$ lie on the  half-line $\overrightarrow{OB_i}$, and $\overleftrightarrow{A_i'B_i'}$, $\overleftrightarrow{A_iB_i}$ and $\overleftrightarrow{a_i'b_i'}$ are parallel to each other.

 For choosing Voronoi centers, we construct $\xi$-distanced equi-spaced grids at each edge of $R$ and use the induced grid points  in the interior of $R$ associated with $d$-dimensional rectangles with edge length $\xi$. The number of centers that fall in the  rectangle $R$ with edge length $L_i$ is of the order of $(L/\xi)^d$. Let, ${\bf c}_1, \dots, {\bf c}_{M_0}$ be the centers  that fall within  $R_{(1)}$. Let $R_1,\dots,R_{M_0}$, be the corresponding Voronoi regions. 
 
 The proposed Voronoi tessellation has centers at  ${\bf c}_1, \dots, {\bf c}_{M_0}$ and at grid-centers that lie on the sides of $R$.  Note that for any point in $R_{(1)}$, the corresponding Voronoi center cannot be outside $R_{(1)}$. For any such point, the Voronoi centers that lie on the sides of $R$ have at least $2\xi d$ distance. At the same time,  for any point in $R_{(1)}$, that  point lies in a  $\xi$  edge length $d$-dimensional rectangular grid, where its vertices are in  the set $\{{\bf c}_1, \dots, {\bf c}_{M_0}\}$. Hence, we have a vertex  at maximum $\xi d$ distance (a very conservative bound). Let, $\{{\bf C}\}=\{{\bf c}_1, \dots, {\bf c}_{M_0}\}$ and let  $ \{\tilde{\bf C}\}$  be the union of $\{{\bf C}\}$ and the grid-centers that lie on the sides of $R$.
 
 Similarly, for  any point outside $R$, the corresponding Voronoi center has to be outside of $R_{(1)}$. We use  regions corresponding to $\{{\bf C}\}$ to approximate $R$. 
 
 Note $ \cup_{i=1}^{M_0}R_i=R^*\subset R$ and $\mathscr{L} (R^* \Delta R) \leq \mathscr{L} (R \Delta R_{(1)} )\leq 8d\xi L$. Here, $\xi$ can be arbitrarily small, thus proving our claim.

 \subsection{Proof of Proposition \ref{vor2}}
 From the proof of Proposition \ref{vor1}, we shift  each center  $\bf {c}_1,\dots,\bf{c}_{M_0}$ by less than or equal to $\delta$ distance in  Euclidean norm.
 
 
 After the perturbation let ${\bf c}_i^*$, $i=1,\dots,M_0$ be the new Voronoi centers. Then, for $\delta<.5\xi$, and for any point in $R_{(1)}$, the corresponding Voronoi centers must be in ${\bf c}^*_1,\dots,{\bf c}^*_{M_0}$ as the distance between cannot be more than $ d\xi+.5\xi$, which follows from the following argument.
 
 As earlier in Proposition \ref{vor1}, points in $R_{(1)}$ will be  inside an $\xi$ edge length d-dimensional rectangle induced by the grid construction.  Therefore, points in $R_{(1)}$ will have a vertex  from the set  $\{{\bf c}_1,\dots,{\bf c}_{M_0}\}$ within $\xi d$ distance. Also, there exists
 ${\bf c}^*_i$'s within $.5\xi$ distance from the grid vertices in $R_{(1)}$. Hence, from the triangle inequality,  for any point in $R_{(1)}$, there exists a point in  $\{{\bf c}^*_1,\dots,{\bf c}^*_{M_0}\}$ within distance $ d\xi+.5\xi$.
 
 From any grid point on the sides of $R$, the distance of any point in $R_{(1)}$ is at least $2d\xi$ and hence, after   perturbing these  grid points by $\delta$, we have a distance of at least $2d\xi-\delta>d\xi+.5\xi$ as $d \geq1$ and $\delta<.5\xi$. Therefore, points in $R_{(1)}$ will be in Voronoi regions corresponding to the  centers  ${\bf c}^*_1,\dots,{\bf c}^*_{M_0}$.

 Similarly, for any points outside $R^{(1)}$, the Voronoi centers cannot be one of $\bf{c}^*_1,\dots,\bf{c}^*_{M_0}$. As, $\mathscr{L} (R^{(1)} \Delta R_{(1)} )$ $\leq 16d\xi L$, it concludes our claim. 
 
 \subsection{General regions}
 We have shown the result for a rectangular region, that is, the covariate space is  partitioned into a  rectangle and its outside. For general regions, the proof follows from writing each partition as a union of non-intersecting finitely many rectangles.

 Suppose, each partition of the covariate space can be written as a non-intersecting union of finitely many rectangles. Let $R^*_1,\dots, R^*_M$ be the $M$ generating rectangles. Let $\{\tilde{\bf C}_i\}$ be the Voronoi centers constructed for $R^*_i$, as in the proof of Proposition \ref{vor1} (using the grid points on the interior rectangle and the grid points that lie on the side of the given rectangle $R_i^*$). Then, $\cup_i \{\tilde{\bf C}_i\}$ gives the Voronoi tessellation for the  covariate space, where each partition element can be approximated by a union of Voronoi regions, with approximation error $O(\xi)$ as in Proposition \ref{vor1}.
 
 For unions of rectangles, we use the union of regions corresponding to Voronoi centers $\cup_{i\in I} \{{\bf C}_i\}$ to approximate 
 $\cup_{i\in{\bf  I}}R^*_i$, for any index set ${\bf I}$, where the approximation error for each $R^*_i$ is $O(\xi)$. Hence, the approximation error is of the order $O(\xi)$ for  $\cup_{i\in{\bf  I}}R^*_i$ , for any ${\bf  I}$. Therefore, the Lebesgue measure of the  Voronoi region with respect to ${\bf B}_c=\cup_{i\in I} \{\tilde{\bf C}_i\} \backslash \cup_{i\in I} \{{\bf C}_i\}$, that is the centers on  the boundaries  of the rectangles,  is  $O(\xi)$. Hence, if $\Omega=\cup_{i=1}^kV_i$, where, $V_l=\cup_{i\in{\bf  I}_l}R^*_i$ for disjoint partitions of ${ I}_l$ of $\{1,\dots,k\}$. Then regions with respect to $\cup_{i\in I_l} \{{\bf C}_i\}$'s are used to approximate $V_l$'s; $l<k$ and  regions corresponding to  $\cup_{i\in I_k} \{{\bf C}_i\}\cup {\bf B}_c $ approximate $V_k$.

 \subsection{Proof of Proposition \ref{vor3}}
 This claim follows as the covariate distribution has positive density over the underlying domain. Therefore, relative frequency of the observations in any open set converges almost surely to the corresponding measure of the open set under measure $\tilde{H}(\cdot)$. Therefore, each small neighborhood around ${\bf c_j}$ has some observation from the covariate vector with probability one,  as $n$ the number of observations, goes to infinity.
 
 In particular let $\tilde{O}$ be any open set and suppose $\tilde{H}(\tilde{O})=\xi'>0$, where $\tilde{H}$ is the underlying measure for the $d$-dimensional covariate distribution. Then,
 $p_i=P(\text{ there is no observation in  }\tilde{O} \text{ for } {\bf x}_i;\ i=1,\dots,n)=(1-\xi')^n$.
 
 As, $\sum p_i<\infty$, then by the Borel-Cantelli lemma, with probability one, there are observations in $\tilde{O}$ for all but finitely many $n$'s. 
 
 \subsection{Proof of Theorem \ref{thm1}}
 
\subsubsection{Sieve Construction} 
 
 Let $\eta(\cdot)$ be any Gaussian process path and $\eta_j(\cdot)$ be corresponding to the  $j$th Voronoi region. For  any $j$th Voronoi region, let $\mathbf{K}^n_j$=$\{\|D^l(\eta_j(.))\|_\infty<M_n,l\leq \alpha, \sigma_i<\lambda_n,\frac{1}{l_i}<\nu_n\}$ and  $\mathbf{K}_{\beta_j}^n=\mathbf{K}_b^n=\{\|\mathbf{\beta}_j\|_{\infty}<\sqrt{n}\}$, where $D^l$ implies the $l$th derivative. Then the corresponding log covering number of smooth Gaussian process paths under supremum norm is $ log(N(\epsilon,\mathbf{K}_j^n,\|\|_\infty))=o(n),$ and  $log(\Pi((\mathbf{K}^n_j)^c))\leq O(-n)$ and $log(\Pi((\mathbf{K}_b)^c))=O(-n)$ (see \cite{tokdar2007posterior};  \citet{ghosal2006posterior}, Lemma 2).
 
 Also, covering  $\mathbf{K}_b^n$ with an $\epsilon$ space grid, the covering number is a polynomial of $n$ of order 1, and $ log(N(\epsilon,\mathbf{K}_b^n,\|\|_\infty))=o(n)$.  
 
 Let, $\mathbf{K}^n_{j,\beta}$ be the prior product space of 
  $\mathbf{K}_{\beta_j}^n$ and $\mathbf{K}^n_j$.  Covering the product space by the product rectangular set $\epsilon$-width grids of  $\mathbf{K}_{\beta}^n$ and $\mathbf{K}^n_j$, from the corresponding spaces, we have the covering number of the product space as the product of two covering numbers.  Hence,  we have   \[log(N(\epsilon,\mathbf{K}_{j,\beta}^n,\|\|_\infty))=log(N(\epsilon,\mathbf{K}_{j}^n,\|\|_\infty))+log(N(\epsilon,\mathbf{K}_b^n,\|\|_\infty))=o(n).\] Also, 
 $log(\Pi((\mathbf{K}^n_{j,\beta})^c))=O(-n)$.
 
 \subsubsection{Entropy bound for the Voronoi tessellation}
 If for the $j$th tessellation, for the coefficients $\|\mathbf{\beta}_{1,j} - \mathbf{\beta}_{2,j}\|_\infty<\epsilon$,   then for the corresponding mean functions,  $\|\mu_{j,1} - \mu_{j,2}\|_\infty<\gamma_1\epsilon$ for some $\gamma_1>0$, as $y$'s are supported on bounded regions. Hence, if $\|\mathbf{\beta}_{1,j} - \mathbf{\beta}_{2,j}\|_\infty<\epsilon$ and $\|\eta_{j,1}-\eta_{j,2}\|_\infty<\epsilon$, then for the corresponding densities $f_1$ and $f_2$, we have $KL(f_1,f_2)<\gamma_2\epsilon$, $\int |f_1-f_2|dy<\gamma_2\epsilon$;  $\gamma_2>0$ (Proposition \ref{neighbor1}).  Choosing, $\alpha\epsilon$ instead of $\epsilon$, with sufficiently small $\alpha>0$,  we have $\int |f_1-f_2|dy<\epsilon$. Hence, writing in terms of $L_1$ covering number,

 \begin{eqnarray}
 log(\Pi((\mathbf{K}^n_{j,\beta})^c) &=& O(-n) \\  log(N(\epsilon,\mathbf{K}^n_{j,\beta},\|\|_1)) &=& o(n).
 \label{covering1} 
 \end{eqnarray}

 Let, $\mathscr{K}_m$, $m<M_{max}$, be the  m-dimensional product space for some Voronoi tessellation, $\mathscr{K}_m=\prod_{l=1}^m \mathbf{K}^n_{l,\beta}$, corresponding to a Voronoi tessellation based on $m$ centers.  
 Let $\mathscr{K}$ be the union of all such  ${n \choose m},\ m\leq M_{max}$, many combinations of $\mathscr{K}_m$'s for all possible Voronoi center selection. As, $log ({n \choose m}) =o(n)$, then using equation \eqref{covering1}, we have:
 \begin{eqnarray}
 log(\Pi(\mathscr{K}^c))=O(-n+log n)=O(-n) \nonumber \\ 
 log(N(\epsilon,\mathscr{K},\|\|_1))=o(n+logn)=o(n),
 \label{cnd}
 \end{eqnarray}
 where $N(\epsilon,\mathscr{K},\|\|_1)$ is the $L_1$ covering number of $\mathscr{K}$. 
 
 %
 \subsubsection{Combining the parts}
 Let $U_{\epsilon'}$ be an $\epsilon'$ radius $L_1$ ball around $f^*$. Then,
 \[\Pi(U_{\epsilon'}^c \mid \cdot)\leq \frac{\int_{U_{\epsilon'}^c}e^{-log(f^*(\mathbf{y}))+ log(f(\mathbf{y}))} d\Pi()}  {(1/M_{max})\int_{N_{1,2}^{\epsilon_2}}e^{-log(f^*(\mathbf{y}))+ log(f(\mathbf{y}))-Mlog(n)} d\Pi()  } =\frac{{\bf N}^n_1}{{\bf D}^n_1}.\]
 
 Here, $\Pi$ denotes the prior distribution and ${\bf y}$ denotes the $n$ length observation vector.  Let, $ N_{1,2}^{\epsilon_2}$ be the set $\eta_j \in N_1$ and $\mu_j \in  N_2$, from Proposition \ref{neighbor1}. For any fixed $\alpha_1,\epsilon_3>0$, choosing $\epsilon_2$ small enough,  for $\eta_j()$ and $\bf{\beta}_j$  in $N_1$ and $N_2$, and for the corresponding density $f$, we have $KL(f^*,f)<\alpha_1\epsilon_3$. 
 Also, $\Pi(N_{1,2}^{\epsilon_2})>0$. (\cite{tokdar2007posterior}) 
 
 This is similar to settings of \cite{tokdar2007posterior}, other than an extra prior term  $e^{-Mlogn}$ in the denominator corresponding to the Voronoi center selection probability, which goes to zero at a polynomial rate. Note that,  $e^{n\alpha_1\epsilon_3}e^{-Mlogn}>1$ as $n$ goes to infinity for any $\alpha_1>0$. Hence, the proof follows. We give a brief sketch in the following argument.  See \citet{ghoshnonparametric} (Theorem 4.4.3 and  4.4.4 proof) for the details and related test construction that we use next.

 Writing, 
 
 \[\frac{{\bf N}^n_1}{{\bf D}^n_1}=e^{2n\alpha_1 \epsilon_3}\frac{{{\bf N}^n_1}_{\mathscr{K}}+{{\bf N}^n_1}_{\mathscr{K}^c}}{e^{2n\alpha_1\epsilon_3}{\bf D}^n_1}\]
 where the subscript ${\mathscr{K}}$ denotes the integral over $U_{\epsilon'}^c \cap {\mathscr{K}}$, and subscript ${\mathscr{K}}^c$ denotes the integral over $U_\epsilon'^c \cap {\mathscr{K}^c}$.

 We can construct test functions $0\leq\Psi_n\leq1$ such that $E_{f^*}(\Psi_n)<e^{-c_1n{\epsilon'}^2/8}$ and $\sup_{f \in U_{\epsilon'}^c}E_f(1-\Psi_n)<e^{-c_1n{\epsilon'}^2/8}$, for any $c_1<1$, for large $n$. We write, 
 \[\Pi(U_{\epsilon'}^c \mid \cdot)=\Psi_n\Pi(U_{\epsilon'}^c \mid \cdot)+(1-\Psi_n)\Pi(U_{\epsilon'}^c \mid \cdot)\leq \Psi_n\Pi(U_{\epsilon'}^c \mid \cdot)+(1-\Psi_n)\frac{{\bf N}^n_1}{{\bf D}^n_1}.\]
 Then, $E_{f^*}((1-\Psi_n){{\bf N}^n_1}_{\mathscr{K}})\leq \sup_{f \in U_{\epsilon'}^c}E_f(1-\Psi_n)<e^{-c_1n{\epsilon'}^2/8}$ and  $E_{f^*}[(1-\Psi_n){{\bf N}^n_1}_{\mathscr{K}^c}]\leq E_{f^*}({{\bf N}^n_1}_{\mathscr{K}^c})<e^{-c_2n}$, $c_2>0$.  
 
 We have, $\Pi(N_{1,2}^{\epsilon_2})>0$ and in $N_{1,2}^{\epsilon_2}$  the integrand  in  ${{\bf D}^n_1}$ is greater than $e^{-2n\alpha_1\epsilon_3}$ with probability one.
 Hence, choosing $\epsilon_3={\epsilon'}^2$, $2\alpha_1<c_1/16$,  \[E(\Pi(U_{\epsilon'}^c \mid \cdot))\leq e^{-c_1n{\epsilon'}^2/8}+e^{-c_1n{\epsilon'}^2/16}.\]
 Using Markov inequality and Borel-Cantelli Lemma, as $P(\Pi(U_{\epsilon'}^c \mid \cdot )>e^{-c_1n{\epsilon'}^2/32}$ $\text{infinitely often})$$\leq \lim _{k\rightarrow \infty}\sum_k^\infty 2 e^{-c_1n{\epsilon'}^2/32}\rightarrow 0$ and  $\Pi(U_{\epsilon'}^c \mid \cdot)$ goes to zero almost surely. 
 
 %
 
 \subsection{General weight function for the norm}
 
 \subsubsection{Voronoi centers derivation}
 Let $\bf{w}^*$ be the true $d$-dimensional  weight vector with entries $w_i^*>0$. Let ${\bf x}_w^*=T_{w^*}({\bf x})$ be the transformed $d$-dimensional scaled covariate vector with $x_{i,w}^*={{w}_i^*}^{\frac{1}{2}}x_i$. We can create the Voronoi tessellation given in Proposition \ref{vor1} and extend it to a general region of a transformed covariate space ${\bf x}_w^*$ and create the partition, as Euclidean distance in ${\bf x}^*$ is equivalent to the weighted metric. Therefore, centers in the covariate space can be achieved by performing an inverse transformation on the centers achieved by this scaling. 
 
 Let $\delta_w$ be a small constant such that for small $\delta_w$ neighborhood of ${\bf w}^*$, that is $\|{\bf w}-{\bf w}^*\|_\infty<\delta_w,  w_i^*>\delta_w \ \forall i$. Let,  ${\bf x}_1$ and ${\bf x}_2$ be two points in the covariate space,  and   ${\bf x}^*_1$ and ${\bf x}^*_2$, and ${\bf x}_{1,w}$ and ${\bf x}_{2,w}$ be their transformed versions for weights ${\bf w}^*$ and ${\bf w}$. 
 
Let $d_w({\bf x}_1,{\bf x}_2)$ be the metric for weight ${\bf w}$.  Choosing $\delta_w$ small enough, we can have $|d_w({\bf x}_1,{\bf x}_2)-d_{w^*}({\bf x}_1,{\bf x}_2)|<\delta$ if $|w_i-w_i^*|<\delta_w, \forall i$. 
 
 Choosing $\delta_w$ small, we can have for ${\bf A}\subset \Omega$ :  $\mathscr{L}({\bf A})<\epsilon\implies \mathscr{L}(T_w({\bf A}))<k_1\epsilon$, and $\mathscr{L}(T_w({\bf A}))<\epsilon\implies \mathscr{L}({\bf A})<k_1\epsilon$; for some  constant $k_1>0$ for the set  $ \{{\bf w}:\|{\bf w}-{\bf w}*\|_\infty<\delta_w\}$. This step follows from a change of variable argument using the fact  $w_i$'s are uniformly  bounded away from zero on a $\delta_w$ neighborhood of ${\bf w}^*$.
 
 Let ${\bf c}_1^{w^*},\dots,{\bf c}_M^{w^*}$ be the Voronoi centers in the covariate space under metric $d_{w^*}$ and with regions $R_1,\dots,R_M$ for the setting of Proposition \ref{vor1}. Therefore, under metric $d_{w^*}$, for  true ${\bf w}^*$, we have unions of regions,  $U_l$'s, approximating the  true partition $V_l$'s. 
 
 Let, $S_{\delta,w}=\{{\bf x} \in \Omega: |d_w({\bf x},{\bf c}_i)-d_w({\bf x},{\bf c}_j)|<2\delta \text { for some } i,j\}$. Letting $\delta$ decrease to zero,  $S_{\delta,w}$ decreases to a   set  formed by  $M \choose 2$ many $d-1$ dimensional  hyperplanes which have equal distances from any two Voronoi centers. Each of the hyperplanes have measure zero. Hence, for any $\delta_1>0$, we can  choose $\delta$ small enough such that $\mathscr{L}(S_{\delta,w^*})<\delta_1$.
 
 Suppose  for  any ${\bf w}$ in the  $\delta_w$ neighborhood of ${\bf w}^*$,  we have ${\bf x} \in \Omega$ in the Voronoi region corresponding to ${\bf c}_i^{w^*}$ under $d_{w^*}$. Then we show that  if ${\bf x} \notin S_{\delta,w^*}$, ${\bf x}$ lies  in the Voronoi region with the same center under metric $d_{w}$ and vice versa. 
 
 The last step follows from the fact that for any point in the covariate space the distances from the centers under two metrics $d_w$ and $d_{w^*}$ differ by at most $\delta$; hence, if $|d_{w^*}({\bf x},{\bf c}_i)-d_{w^*}({\bf x},{\bf c}_j)|>2\delta$ and $d_{w^*}({\bf x},{\bf c}_i)<d_{w^*}({\bf x},{\bf c}_j)$, then $d_{w}({\bf x},{\bf c}_i)<d_{w}({\bf x},{\bf c}_j)$. 
 
 Let $\xi>0$ be any small constant. For the centers ${\bf c}_1^{w^*},\dots,{\bf c}_M^{w^*}$,  we have  $R_1,\dots,R_M$ the  Voronoi region under $d_{w^*}$.  Then, choosing $\delta_w$ small,  for  $\tilde{R}_1,\dots,\tilde{R}_M$  the regions under  $d_w$,   we have $\sum R_i\Delta \tilde{R}_i<\xi$, as we can make the measure of the set $S_{\delta,w^*}$ arbitrary small by choosing small $\delta_w$. 
 
 Suppose,  we have $U_l$ and $\tilde{U}_l$'s corresponding to $d_{w^*}$ and $d_w$, respectively. Then, $\sum V_i\Delta \tilde{U}_i<\sum  V_i\Delta {U}_i+\xi$. Hence, choosing $\xi$ small enough,  we can use the ${\bf w}$ in the $\delta_w$ neighborhood of the  true weight  ${\bf w^*}$ to approximate the covariate space  partition using Voronoi regions.

 %
 %
 
 \subsubsection{Prior mass condition}
 Hence, we can use the Voronoi centers for ${\bf w}^*$ for a $\delta_w$ supremum neighborhood around ${\bf w}^*$ for approximating the true partition of the parameter space. Under Dirichlet a prior, that  neighborhood has positive probability and hence the  proof of Theorem \ref{thm1} holds. 
 %
 %
 %
 \subsection{Special case: true partition is a Voronoi   tessellation }
 Suppose the  true partition of the covariate space is given by $M$ Voronoi regions corresponding to centers ${\bf c}_1^*,\dots, {\bf c}_M^*$. Let, ${\bf c}_1,\dots, {\bf c}_M$ be in the neighborhood of ${\bf c}_i^*$  with $d({\bf c}_i,{\bf c}_i^*)<\delta_c$. Then, for any ${\bf x}\in \Omega$,  $|d({\bf x},{\bf c}_i)-d({\bf x},{\bf c}_i^*)|<\delta_c$. 
 
 Let $S_{\delta_c}=\{{\bf x} \in \Omega: |d({\bf x},{\bf c}_i^*)-d({\bf x},{\bf c}_j^*)|<2\delta_c \text { for some } i,j\}$. We choose $\delta_1>0$ to be any small constant. Choosing $\delta_c$ small enough,  we can have $\mathscr{L}(S_{\delta_c})<\delta_1$ from the argument given in Section 7.6.1  for the general weight function. Outside $S_{\delta_c}$, we have $d({\bf x},{\bf c}_i^*)<d({\bf x},{\bf c}_j^*) \implies d({\bf x},{\bf c}_i)<d({\bf x},{\bf c}_j)$ and vice versa (triangle inequality). Therefore, for $R_1,\dots,R_M$ the regions corresponding to ${\bf c}_1^*,\dots, {\bf c}_M^*$ and $\tilde{R}_1,\dots,\tilde{R}_M$ the regions corresponding to ${\bf c}_1,\dots, {\bf c}_M$, we have $\sum R_i\Delta \tilde{R}_i<\delta_1$. 
 
 Hence, by picking centers at  $\delta_c$ neighborhoods around the true Voronoi centers, we can approximate the   region with  $\delta_1$ accuracy. From Proposition \ref{vor3}, we  have  observations in $\delta_c$ neighborhood with probability 1 and can choose our Voronoi tessellation.

 \subsection{Proof of Proposition \ref {neighbor1}}
 This proof follows using techniques similar to  \cite{tokdar2007posterior}. A sketch can be  given as follows.  The set where the two Gaussian process paths are more than $\epsilon$ away in $N_1$ has Lebesgue measure of order $\epsilon$ (by the construction of $N_1$ with $\epsilon_2=\epsilon_1=\epsilon$) and hence, its probability under $\tilde{H}$ is less than  $k_3\epsilon$ where $k_3>0$ is fixed (depends on the bounded density function $\tilde{h}$ of $\tilde{H}(.)$).   Consider $f_1$ and $f_2$ with corresponding ${\bf \beta}_1$, ${\bf \beta}_2$, $\eta_1(\cdot)$ and $\eta_2(\cdot)$ in $N_1$ and $ N_2$, respectively.  Then, for the  distribution corresponding to $f_1$ and $f_2$, $|g_{\bf x}^{f_1}(y)-g_{\bf x}^{f_2}(y)|=g_{\bf x}^{f_1}(y)O(\epsilon)$ outside of  an $O(\epsilon)$ measure set of  the covariate space.   Then, from the fact that $y$ is supported on a bounded region, $ |\int g_{\bf x}^{f_1}(y)d(x,y)-\int g_{\bf x}^{f_2}(y)d(x,y)|=O(\epsilon) $ and  the proof follows. 
 
\bibliographystyle{abbrvnat}
\bibliography{references}

\end{document}